\documentclass[twocolumn,tighten]{aastex63}
\usepackage{CJKutf8}
\usepackage{times}
\usepackage{amsmath}
\usepackage{graphicx}
\usepackage{subfigure}
\usepackage{placeins}
\usepackage{hyperref}
\usepackage{gensymb}
\usepackage{upgreek}
\newcommand{\kms}{km\ s$^{\mathrm{-1}}$}
\newcommand{\HI}{H{\sevenrm\,I}}
\newcommand{\MHI}{$M_{\mathrm{H\, \textsc{i}}}$}
\newcommand{\mstar}{$M_{*}$}
\newcommand{\msun}{$M_{\mathrm{\odot}}$}
\newcommand{\sersic}{S\'ersic}

 \font\sevenrm=cmr7 scaled 1000

\received{\today}
\revised{xxxx , 2022}
\accepted{xxxx , 2022}
\submitjournal{ApJ}

\begin{document}
\begin{CJK*}{UTF8}{gbsn} 

\title{Centrally Concentrated \HI\, Distribution Enhances Star Formation in Galaxies}

\correspondingauthor{Jing Wang}
\email{jwang\_astro@pku.edu.cn}

\shorttitle{Influence of Profile Shape and Asymmetry on Star Formation}
\shortauthors{YU, WANG \& HO}

\author[0000-0002-9066-370X]{Niankun Yu (余捻坤)}
\affiliation{Kavli Institute for Astronomy and Astrophysics, Peking University, Beijing 100871, China}
\affiliation{Department of Astronomy, School of Physics, Peking University, Beijing 100871, China}

\author[0000-0001-6947-5846]{Luis C. Ho}
\affiliation{Kavli Institute for Astronomy and Astrophysics, Peking University, Beijing 100871, China}
\affiliation{Department of Astronomy, School of Physics, Peking University, Beijing 100871, China}

\author[0000-0002-6593-8820]{Jing Wang}
\affiliation{Kavli Institute for Astronomy and Astrophysics, Peking University, Beijing 100871, China}
\affiliation{Department of Astronomy, School of Physics, Peking University, Beijing 100871, China}

\begin{abstract}
We use a sample of 13,511 nearby galaxies from the ALFALFA and SDSS spectroscopic surveys to study the relation between the spatial distribution of \HI\ 21~cm emission and star formation rate (SFR).  We introduce a new non-parametric quantity $K$, measured from the curve-of-growth of the line, to describe the shape of the integrated \HI\ profile. The value of $K$ increases from double-horned to single-peaked profiles, depending on projection effects and the spatial and velocity distribution of the gas. Using carefully chosen samples to control for the competing factors that influence the integrated line profile, we argue that useful inferences can be made on the spatial distribution of the gas.  We find that galaxies with a high value of $K$ tend to have more centrally concentrated \HI\ distribution within the optical disk of the galaxy at fixed conditions, and that larger values of $K$ are associated with higher levels of total and central SFR.  The results suggest that the global concentration of \HI\ plays an important role in facilitating the conversion of neutral atomic hydrogen to molecular hydrogen gas, which, in turn, affects the star formation activity throughout the optical disk. Our sample is biased against quiescent galaxies, and thus the conclusions may not hold for galaxies with low SFR or low \HI\ content.
\end{abstract}


\section{Introduction} \label{sec:intro}

Star formation is a principal process of galaxy formation and evolution (\citealt{Kennicutt2012ARA&A..50..531K}, and references therein). A tight relation between the star formation rate (SFR) and stellar mass (\mstar) is found for star-forming galaxies at different redshifts (e.g., \citealt{Noeske2007ApJ...660L..43N, Elbaz2011A&A...533A.119E, Tacchella2018ApJ...859...56T, Tacconi2018ApJ...853..179T}), which can be reproduced by cosmological simulations (e.g., \citealt{Sparre2015MNRAS.447.3548S}). This tight relation defines the star-forming galaxy main sequence with a typical scatter of $\sim$0.2--0.4 dex (e.g., \citealt{Rodighiero2011ApJ...739L..40R, Sargent2012ApJ...747L..31S, Whitaker2014ApJ...795..104W}). 

The gas compaction scenario, originally proposed to explain the transition between blue and red nugget galaxies at high redshifts \citep{Damjanov2009ApJ...695..101D, Barro2013ApJ...765..104B}, envisions that galaxies oscillate around the main sequence as a consequence of star formation being regulated by gas depletion and accretion (\citealt{Dekel2014MNRAS.438.1870D, Tacchella2016MNRAS.457.2790T}).  The observed scatter of the main sequence reflects the modulation of the SFR and star formation efficiency resulting from the combined effects of strong gas inflows, energy feedback, and gas consumption \citep{Tacchella2016MNRAS.458..242T}.  As many key physical processes are common for the evolution of low-redshift and high-redshift galaxies \citep{Somerville2015ARA&A..53...51S}, the compaction scenario can be generalized to account for the secular evolution of low-redshift galaxies, whose characteristically lower accretion rates and stable disks result in weaker, more prolonged compaction cycles \citep{Zolotov2015MNRAS.450.2327Z, Tacchella2016MNRAS.457.2790T}.  At low redshifts, gas inflows are facilitated by non-axisymmetric perturbations to the gravitational potential induced by bars, spiral arms, and lopsidedness \citep{Hernquist1989Natur.340..687H, Regan2004ApJ...600..595R, Hopkins2011MNRAS.415.1027H}, which accelerate the concentration of the gas and enhance central star formation \citep{KK2004}. The central accumulation of gas and its subsequent formation of stars play a key role in building up stellar mass in the central regions of galaxies (e.g. \citealt{Ellison2018MNRAS.474.2039E, Wang2018ApJ...865...49W, Luo2020MNRAS.493.1686L}). 

Star formation is a localized instead of a global process, and its rate is determined by the local gas density \citep{Bigiel2008AJ....136.2846B, Krumholz2018MNRAS.477.2716K} and regulated by local conditions, including pressure \citep{Kapferer2009A&A...499...87K} and metallicity \citep{Poetrodjojo2018MNRAS.479.5235P}.  The volumetric or surface density of both molecular and atomic gas is vital for regulating star formation.  While molecular hydrogen is more directly related to the star formation process (\citealt{Schmidt1959ApJ...129..243S, Kennicutt1998ARA&A..36..189K, Bigiel2008AJ....136.2846B, Kennicutt2012ARA&A..50..531K}), neutral atomic hydrogen serves as the reservoir of raw material that fuels the production of molecular hydrogen, and eventually stars (e.g., \citealt{Leroy2008AJ....136.2782L, Hess2018AAS...23121005H, Pokhrel2019AAS...23346005P}).  Theoretical models show that an atomic hydrogen surface density of $\sim$ 10 \mstar\ pc$^{-2}$ is required to shield molecular hydrogen against dissociation, and the molecular fraction in a galaxy is determined by its density and metallicity \citep{Krumholz2009ApJ...693..216K}.  On average, central galaxies are found depleted of their \HI\, reservoir once they drop off the star-forming main sequence (\citealt{Cortese2020MNRAS.494L..42C}; \citealt{Guo2021ApJ...918...53G}; but see different results in \citealt{Zhang2019ApJ...884L..52Z} for massive spiral galaxies).  The specific SFR (sSFR $\equiv$ SFR/$M_*$) rises with increasing \HI\, mass fraction (\citealt{Doyle2006MNRAS.372..977D, Lah2007MNRAS.376.1357L, Huang2012ApJ...756..113H, Saintonge2016MNRAS.462.1749S, Saintonge2018MNRAS.481.3497S, Zhou2018PASP..130i4101Z}), even though the scatter is relatively large. Therefore, \HI\, is a key component for star formation and galaxy evolution, and it is highly desirable to know the spatial location of the gas and its distribution relative to the stars in a galaxy.

\begin{figure*}[t]
\centering
\includegraphics[width=1.0\textwidth]{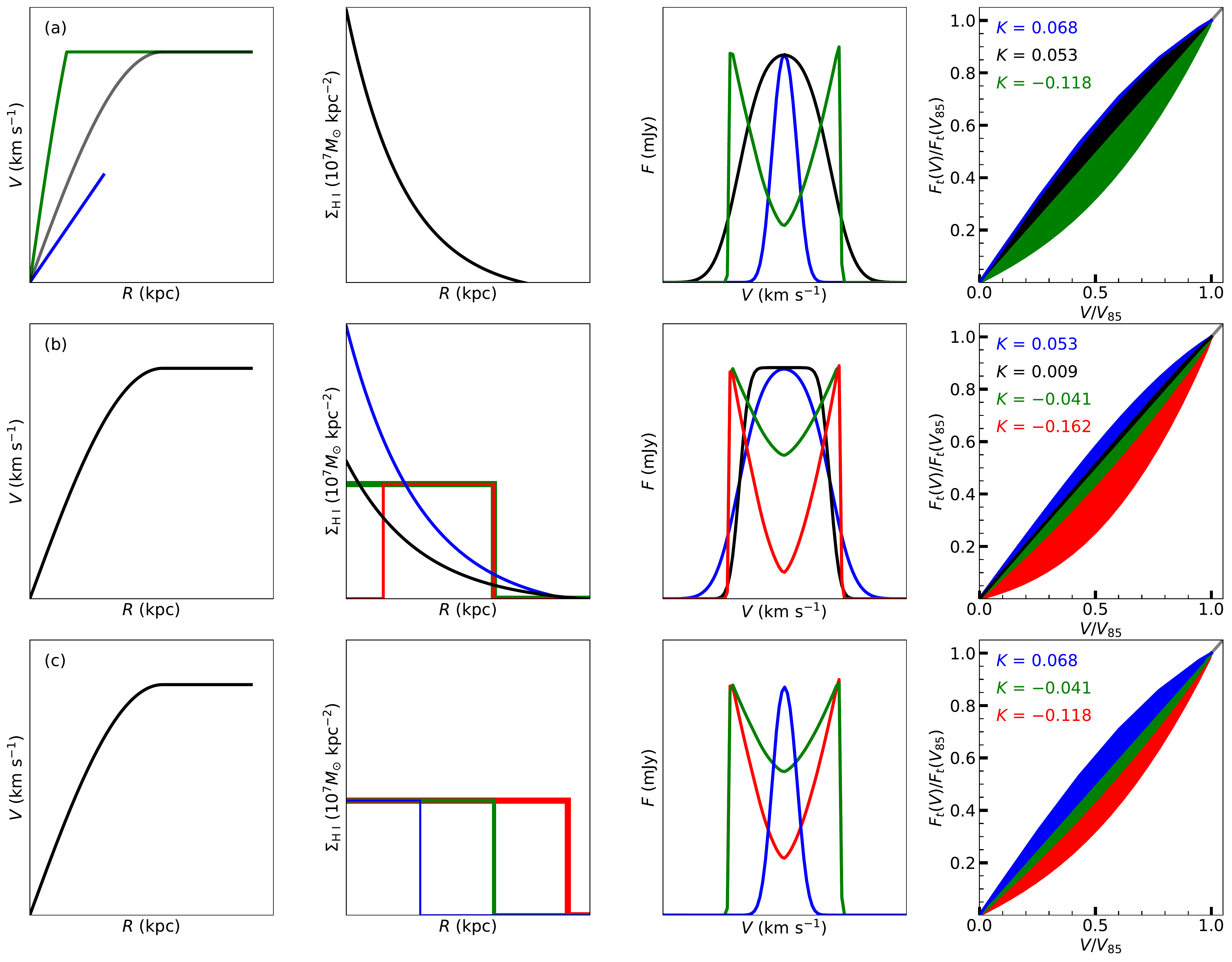}
\caption{Schematic view to illustrate how the (a) rotation curve, (b) \HI\ radial distribution, and (c) \HI\ radius influence the observed \HI\, profile shape and profile shape parameter $K$. From left to right, the panels in each row show the rotation curve, \HI\ distribution, corresponding \HI\, profile, and the normalized curve-of-growth as well as the corresponding value of $K$.  The galaxy inclination angle is assumed to be the same.
}
\label{fig:toyModel}
\end{figure*}

\begin{figure*}[ht]
\epsscale{0.9}
\plotone{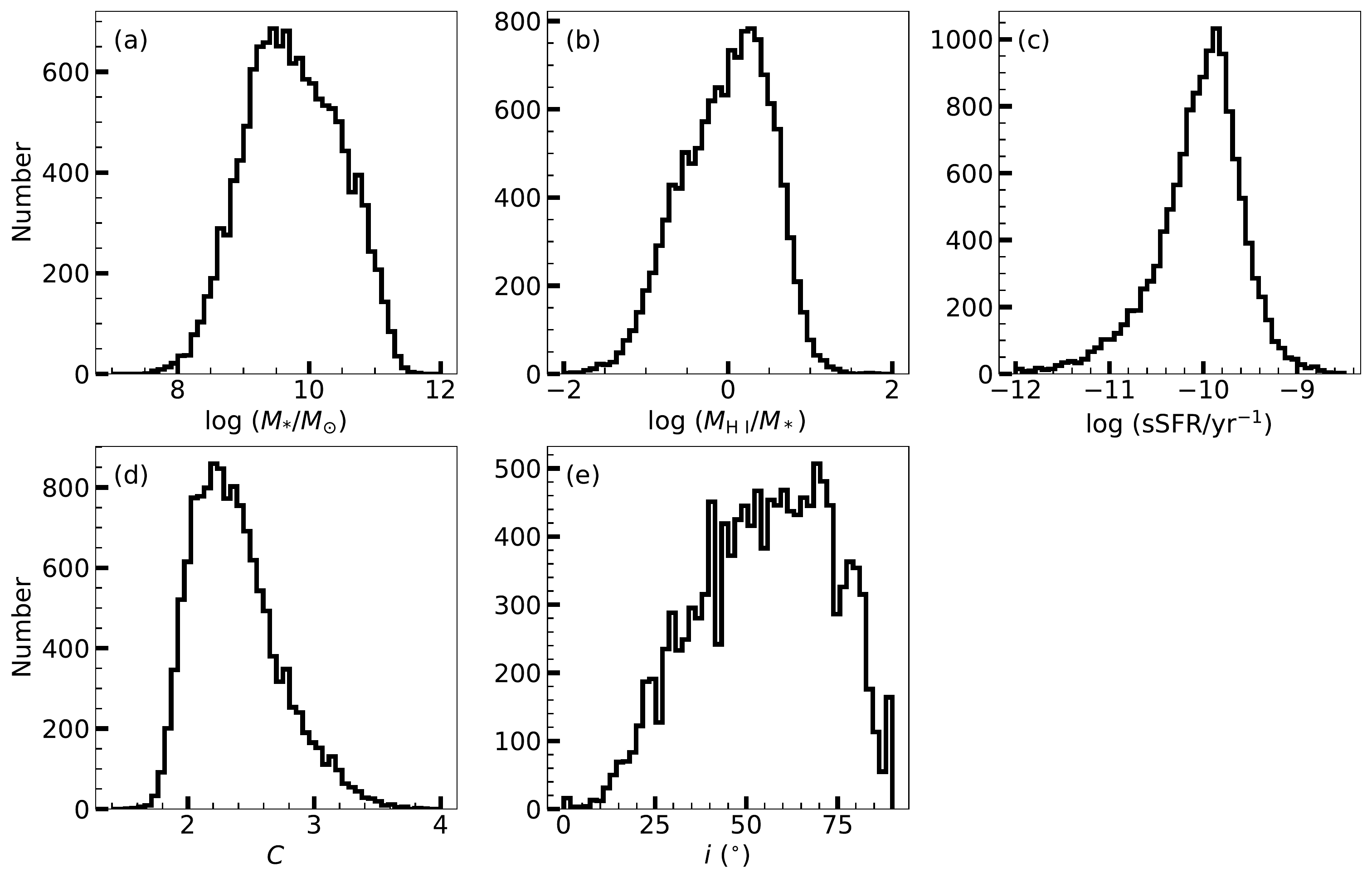}
\caption{Sample distribution of (a) stellar mass, (b) ratio of \HI\, mass to stellar mass, (c) sSFR, (d) optical concentration in the $r$ band, and (e) inclination angle.}
\label{fig:paraTable}
\end{figure*}

Among disk-dominated galaxies, the SFR enhancement with respect to the star-forming main sequence on average scales up with the average \HI\, surface density within the optical disk \citep{Wang2020ApJ...890...63W}. However, apart from nearby dwarf irregular galaxies, whose young stars largely spatially overlap with the neutral atomic hydrogen (\citealt{Hunter2021AJ....161...71H}), in more massive galaxies the \HI\, gas is normally distributed in a much more extended configuration than the optical stellar disk, and its surface density does not increase as quickly as H$_2$ or SFR toward the central region (\citealt{Wong2002ApJ...569..157W, Bigiel2008AJ....136.2846B}). The \HI\, distribution is traditionally derived from spatially resolved maps (e.g., \citealt{Rots1980A&AS...41..189R, Yun1994Natur.372..530Y, Noordermeer2005A&A...442..137N}) or azimuthally averaged radial profiles (e.g., \citealt{Baldwin1980MNRAS.193..313B, RheevanAlbada1996AAS..115..407R, Hunter2012AJ....144..134H}). The \HI\, disk traces the spiral arms and outer disk (e.g., \citealt{Hunter2012AJ....144..134H, Wang2013MNRAS.433..270W}), and often exhibits morphological features such as rings (e.g., \citealt{Emonts2008A&A...488..519E, Bait2020MNRAS.492....1B}), warps (e.g., \citealt{Martinsson2016AA...585A..99M}), and holes created either by local events in the disk (e.g., \citealt{Boomsma2008A&A...490..555B, Warren2011ApJ...738...10W}) or dynamic stability of the central regions of galaxies (e.g., \citealt{Murugeshan2019MNRAS.483.2398M}).  The \HI\, radial profile, when scaled to the \HI\, radius, shows a central depression, followed by an exponential (\citealt{Swaters2002AA...390..829S, Wang2014MNRAS.441.2159W, Wang2016MNRAS.460.2143W}) or S\'ersic (1963) function (\citealt{Hunter2021AJ....161...71H}) further out.
 
Apart from small samples gathered for targeted investigations (e.g., \citealt{Noordermeer2005A&A...442..137N, Walter2008AJ....136.2563W, Martinsson2016AA...585A..99M}), \HI\, maps or radial profiles of galaxies are not widely available for reliable statistical analysis. By contrast, single-dish integrated \HI\, spectra of tens of thousands of galaxies have been accumulated from large-scale survey programs such as the H~I Parkes All-Sky Survey (HIPASS: \citealt{Koribalski2004AJ....128...16K}; \citealt{Meyer2004MNRAS.350.1195M}; \citealt{Wong2006MNRAS.371.1855W}), the GALEX Arecibo SDSS Survey (GASS: \citealt{Catinella2010MNRAS.403..683C, Catinella2012A&A...544A..65C, Catinella2013MNRAS.436...34C}; GASS-low: \citealt{Catinella2018MNRAS.476..875C}), and the Arecibo Legacy Fast ALFA survey (ALFALFA: \citealt{Haynes2011AJ....142..170H, Haynes2018ApJ...861...49H}).  Moreover, upcoming surveys using the Australian Square Kilometer Array Pathfinder (ASKAP: \citealt{Johnston2007PASA...24..174J}), the Five-hundred-meter Aperture Spherical radio Telescope (FAST: \citealt{Nan2011IJMPD..20..989N}), and Apertif \citep{Adams2018AAS...23135404A} will significantly enlarge the \HI\, database for nearby galaxies. 

Although the information content that can be derived from a single-dish spectrum is relatively limited, the 21~cm line, apart from furnishing the recession velocity, neutral atomic hydrogen mass, and velocity width of the galaxy, also provides additional insights on the physical properties of galaxies through its profile {\it shape}.  In relatively isolated, non-interacting galaxies, the observed \HI\ profiles largely fall into three generic categories: single-peaked, flat-topped, and double-horned (e.g., \citealt{Bottinelli+1990AAS...82..391B, Haynes1998AJ....115...62H, Springob2005ApJS..160..149S, Courtois+2009AJ....138.1938C}).  Double-horned profiles are more prevalent in high-mass spiral galaxies than in less massive, late-type systems \citep{Shostak1977A&A....58L..31S,Yu2022}.  Among late-type spirals that do have double-horned \HI\, lines, a minority contain an additional peak at the systematic velocity that may be associated with central gas enhancement due to the effects of a bar \citep{Matthews1998AJ....116.1169M}. In spite of their large masses, ongoing and recent mergers show a preponderance of single-peaked profiles as a consequence of gas inflows induced by gravitational torques \citep{Zuo2022}.

What can be learned from the shape of the integrated \HI\ profile?  After accounting for line-of-sight projection, the observed global line profile primarily reflects the velocity field and spatial distribution of the \HI-emitting gas \citep{El-Badry2018MNRAS.477.1536E}.  Thus, given the known inclination angle---a quantity easy to obtain for nearby galaxies if we assume that the gas lies in a disk that is coplanar with that of the stars---we potentially can extract useful clues on the spatial distribution of the \HI, provided that we can constrain the rotation curve of the galaxy.  In this work, we utilize a new parameter, $K$, introduced in the companion paper by Yu et al. (2022a; see Section~3.1 for more details) and based on the curve-of-growth method of \citet{Yu2020ApJ...898..102Y}, to quantify the shape of the \HI\ profile for a large sample of nearby galaxies derived from ALFALFA and the Sloan Digital Sky Survey Data Release 16 (SDSS DR16: \citealt{AhumadaSDSSDR162020ApJS..249....3A}).  Our method is distinct from previous parameterizations of profile shapes using analytical functions \citep{Stewart2014A&A...567A..61S,WestmeierBF2014MNRAS.438.1176W}, and less subjective than traditional classifications by visual inspection (e.g., \citealt{Matthews1998AJ....116.1169M, Geha2006ApJ...653..240G, Ho2008, Espada2011A&A...532A.117E}).  We use the optical light concentration, which traces the stellar mass distribution and hence the gravitational potential of the galaxy, as a rough proxy for the shape of the rotation curve.  With this information in hand, fixing the inclination angle then allows us to obtain an indirect inference on the radial distribution (concentration) of the gas.  Our goal is to investigate the possible dependence of star formation activity on the spatial distribution of \HI.

Our sample, data, and method are described in Section~\ref{sec:da}.  Section~\ref{sec:HISFR} investigates the dependence of star formation on \HI\ profile shape. Our main conclusions are summarized in Section~\ref{sec:sum}.

\section{Data and Analysis}
\label{sec:da}

\subsection{Sample, \HI\ Measurements, and Other Parameters}
\label{subsec:sample}

We focus on the sample of low-redshift ($z< 0.06$) galaxies contained in the ALFALFA survey that overlaps with SDSS DR16, after excluding objects without a reliable optical counterpart and those contaminated by nearby companions or radio frequency interference (see \citealt{Yu2022} for details). The availability of optical spectroscopic and imaging data is essential, as they furnish several physical parameters critical for our analysis.  As described in \cite{Yu2022}, the galaxy inclination angle ($i$) derives from the $r$-band axis ratio, assuming an intrinsic disk thickness that depends on the stellar mass \citep{Sanchez-Janssen2010MNRAS.406L..65S}.  

We describe the stellar mass distribution using the optical concentration $C \equiv R_{90}/R_{50}$ (\citealt{Strateva2001AJ....122.1861S, Kauffmann2003MNRAS.341...33K}), where $R_{90}$ and $R_{50}$ are the radii enclosing 90\% and 50\% of the $r$-band Petrosian flux, respectively. Other measures of stellar mass concentration can be contemplated, including the \cite{Sersic1963BAAA....6...41S} index $n$, which is available for our sample from \cite{Simard2011ApJS..196...11S}, the stellar mass surface density within the inner 1 kpc (\citealt{Fang2013ApJ...776...63F, Yesuf2020ApJ...889...14Y}), or the surface density within the effective radius (\citealt{Naab2006MNRAS.369..625N}).  For convenience and ease of comparison with the literature, we use the concentration parameter $C$, which is sensitive to the stellar mass distribution out to a relatively large radius. We confirm that substituting $C$ with the \sersic\, index or the stellar surface density within the effective radius does not significantly change our results.

We use the second version of the GALEX-SDSS-WISE Legacy Catalog with the deepest photometry (GSWLC-X2:, \citealt{Salim2018ApJ...859...11S}) to obtain the total stellar masses and global SFRs.  The total stellar masses and SFRs, derived from fits of the ultraviolet-to-optical spectral energy distribution with constraints from infrared photometry, have formal median statistical uncertainties of 0.042 dex and 0.064 dex, respectively. To quantify the strength of star formation in the central region of the galaxy, we utilize SFR$_{\rm in}$ from the MPA-JHU catalog (\citealt{Kauffmann2003MNRAS.341...33K, Brinchmann2004MNRAS.351.1151B}), which pertains to signal collected within the 3\arcsec-diameter SDSS fiber\footnote{At a median distance of 137 Mpc for our sample, 3\arcsec\ corresponds to $\sim 2$ kpc.}. The inner SFR is based on the extinction-corrected H$\alpha$ emission line and $D_{n}4000$, the strength of the 4000~\AA\ break. All SFRs and stellar masses have been scaled to the stellar initial mass function of \cite{Chabrier2003PASP..115..763C}. Many factors can affect the true uncertainties of the stellar masses and SFRs.  Internal extinction presents a perennial problem, even if, in principle, the spectral energy distribution fits of \cite{Salim2018ApJ...859...11S} formally account for extinction and incorporate constraints from infrared data.  Apart from uncertainties associated with the extinction derived from the Balmer decrement, the fiber-based measurements of SFR$_{\rm in}$ are also affected by systematic uncertainties arising from the fiber covering area and the different timescales of star formation probed by $D_{n}4000$ and H$\alpha$ \citep{Brinchmann2004MNRAS.351.1151B}.

\citet{Yu2022} use the method of \citet{Yu2020ApJ...898..102Y}, which integrates the \HI\ intensity as a function of velocity from the line center outward to both the blue and red sides of the profile, to build the curve-of-growth for each \HI\ spectrum from the ALFALFA survey. Besides the median integrated flux of the flat part of the curve-of-growth, velocity widths that capture characteristic percentages of the total flux, and two measures of line asymmetry, \citet{Yu2020ApJ...898..102Y} introduced the profile concentration parameter $C_V \equiv V_{85}/V_{25}$, where $V_{85}$ and $V_{25}$ are the velocity widths enclosing, respectively, 85\% and 25\% of the total flux of the curve-of-growth. Here, we use the new parameter $K$ proposed by \citet{Yu2022} to quantify the profile shape.  Normalizing the velocity axis by $V_{85}$ and the integrated flux axis by 85\% of the total flux [$F_t(V_{85})$], $K$ is defined as the integrated area between the normalized curve-of-growth and the diagonal line in the right panels of Figure~\ref{fig:toyModel}. The parameter $K$, which uses all the data points on the rising part of the curve-of-growth, is superior to $C_V$, which is based only on two measures of line width.  As illustrated in Figure~\ref{fig:toyModel}, a single-peaked profile is characterized by $K>0$, a perfectly rectangular (flat-topped) profile by $K=0$, and double-peaked profiles have $K<0$.  With increasing $K$, the profile changes from double-peaked to single-peaked.

While the curve-of-growth algorithm generally produces reliable outputs for the ALFALFA data without much manual intervention \citep{Yu2022}, there can be circumstances when special treatment is needed. For example, narrow emission profiles that contain fewer than six emission channels or profiles with strong absorption features cannot be measured automatically without first generating a mask or setting the signal range over which to perform the analysis.

\begin{deluxetable*}{ccrrrrrr}[ht]
\centering
\small\addtolength{\tabcolsep}{3pt}
\tabletypesize{\footnotesize}
\tablecolumns{8}
\tablewidth{0pt} 
\tablecaption{The Relation between Star Formation Rate and $K$}
\tablehead{
\colhead{$C$} &
\colhead{$\log M_{\mathrm{H~I}}$} &
\colhead{log $M_*$} &
\colhead{Number} &
\multicolumn2c{$\Delta \log \rm SFR$}&  
\multicolumn2c{$\Delta \log {\rm SFR}_{\rm in}$} 
\\ 
\colhead{} &  
\colhead{(\msun)} & 
\colhead{(\msun)} & 
\colhead{} &
\multicolumn2c{(\msun\, yr$^{-1}$)}&  
\multicolumn2c{(\msun\, yr$^{-1}$)}
\\
\colhead{} &  
\colhead{} & 
\colhead{} & 
\colhead{} &
\colhead{$r$} &
\colhead{$p$} &
\colhead{$r$} &
\colhead{$p$} 
\\
\colhead{(1)} &
\colhead{(2)} &
\colhead{(3)} &
\colhead{(4)} &
\colhead{(5)} &
\colhead{(6)} &
\colhead{(7)} &
\colhead{(8)}  
}
\decimals 
\startdata
 2.0$-$2.3 & 9.3$-$9.6 & 8.5$-$9.1 & 170 & 0.26 & 0.00 & 0.29 & 0.00 \\
 &  & 9.1$-$9.7 & 194 & 0.17 & 0.01 & 0.18 & 0.01 \\
 &  & 9.7$-$10.3 & 61 & 0.45 & 0.00 & 0.35 & 0.01 \\
  \hline
  2.0$-$2.3 & 9.6$-$9.9 & 8.5$-$9.1 & 69 & 0.13 & 0.28 & 0.22 & 0.07 \\
 &  & 9.1$-$9.7 & 265 & 0.20 & 0.00 & 0.16 & 0.01 \\
 &  & 9.7$-$10.3 & 199 & 0.22 & 0.00 & 0.26 & 0.00 \\
 &  & 10.3$-$10.9 & 68 & 0.22 & 0.08 & 0.16 & 0.18 \\
 \hline
  2.0$-$2.3 & 9.9$-$10.2 & 9.1$-$9.7 & 143 & 0.09 & 0.30 & 0.12 & 0.16 \\
 &  & 9.7$-$10.3 & 194 & 0.16 & 0.03 & 0.32 & 0.00 \\
 &  & 10.3$-$10.9 & 123 & 0.20 & 0.03 & 0.29 & 0.00 \\
 \hline
2.3$-$2.6 & 9.3$-$9.6 & 8.5$-$9.1 & 95 & 0.14 & 0.19 & 0.20 & 0.06 \\
 &  & 9.1$-$9.7 & 99 & 0.23 & 0.02 & 0.37 & 0.00 \\
 &  & 9.7$-$10.3 & 48 & 0.48 & 0.00 & 0.53 & 0.00 \\
  \hline
 2.3$-$2.6 & 9.6$-$9.9 & 8.5$-$9.1 & 44 & 0.42 & 0.00 & 0.39 & 0.01 \\
 &  & 9.1$-$9.7 & 159 & 0.33 & 0.00 & 0.37 & 0.00 \\
 &  & 9.7$-$10.3 & 119 & 0.41 & 0.00 & 0.30 & 0.00 \\
 &  & 10.3$-$10.9 & 61 & 0.02 & 0.89 & $-$0.02 & 0.88 \\
  \hline
2.3$-$2.6 & 9.9$-$10.2 & 9.1$-$9.7 & 91 & 0.19 & 0.07 & 0.32 & 0.00 \\
 &  & 9.7$-$10.3 & 160 & 0.34 & 0.00 & 0.33 & 0.00 \\
 &  & 10.3$-$10.9 & 114 & 0.18 & 0.06 & 0.29 & 0.00 \\
  \hline
2.6$-$2.9 & 9.3$-$9.6 & 8.5$-$9.1 & 40 & 0.23 & 0.15 & 0.10 & 0.56 \\
 &  & 9.1$-$9.7 & 37 & 0.08 & 0.64 & 0.10 & 0.57 \\
  \hline
2.6$-$2.9 & 9.6$-$9.9 & 9.1$-$9.7 & 60 & 0.18 & 0.18 & 0.33 & 0.01 \\
 &  & 9.7$-$10.3 & 58 & 0.42 & 0.00 & 0.33 & 0.01 \\
 &  & 10.3$-$10.9 & 50 & $-$0.01 & 0.92 & $-$0.02 & 0.90 \\
  \hline
2.6$-$2.9 & 9.9$-$10.2 & 9.1$-$9.7 & 34 & 0.36 & 0.04 & 0.36 & 0.04 \\
 &  & 9.7$-$10.3 & 58 & 0.13 & 0.32 & 0.26 & 0.05 \\
 &  & 10.3$-$10.9 & 81 & 0.17 & 0.12 & 0.32 & 0.00 \\
\enddata
\tablecomments{Col. (1): Optical concentration. Col. (2): \HI\, mass. Col. (3): Stellar mass. Col. (4): Number of galaxies. Cols. (5)--(8): Pearson correlation coefficient $r$ and the probability $p$ for rejecting the null hypothesis that there is no relation between $K$ and $\Delta \log \rm SFR$, and $K$ and $\Delta \log {\rm SFR}_{\rm in}$, respectively.  The subsamples only include galaxies with $50^{\circ}\leq\ i <\ 70^{\circ}$, and we only consider bins with at least 30 galaxies.} \label{tab:pearson}
\end{deluxetable*}

The final sample used in this work comprises 13,511 sources, as listed in Table~3 of \citet{Yu2022}.  Note that the distances of the individual galaxies in the sample are adopted from \cite{Haynes2018ApJ...861...49H}, and all distance-dependent quantities from the SDSS databases have been scaled to these adopted distances.  Figure~\ref{fig:paraTable} summarizes some basic properties of the sample.  The sample has stellar masses $M_* \approx 10^8 - 10^{11.5}\,M_\odot$ (median value $10^{9.7}\,M_\odot$) and \HI\ gas mass fraction $M_{\rm H~I}/M_* \approx 0.01$ to 30 (median value $\sim 1$), a factor $\sim 3$ higher than more representative samples in the local Universe (e.g., \citealt{Catinella2018MNRAS.476..875C}).  With a median concentration index of $C = 2.3$ and ${\rm sSFR} \approx 10^{-10}\,{\rm yr}^{-1}$, the ALFALFA sample, derived from a relatively shallow survey that only includes \HI\ detections, is incomplete for early-type ($C> 2.6$; \citealt{Kauffmann2003MNRAS.341...54K}) and quiescent (${\rm sSFR}< 10^{-11}\, {\rm yr}^{-1}$; \citealt{Brinchmann2004MNRAS.351.1151B}) galaxies. Instead, it mainly covers star-forming, gas-rich, and late-type systems.

\begin{figure}[!ht]
\centering
\includegraphics[width=0.5\textwidth]{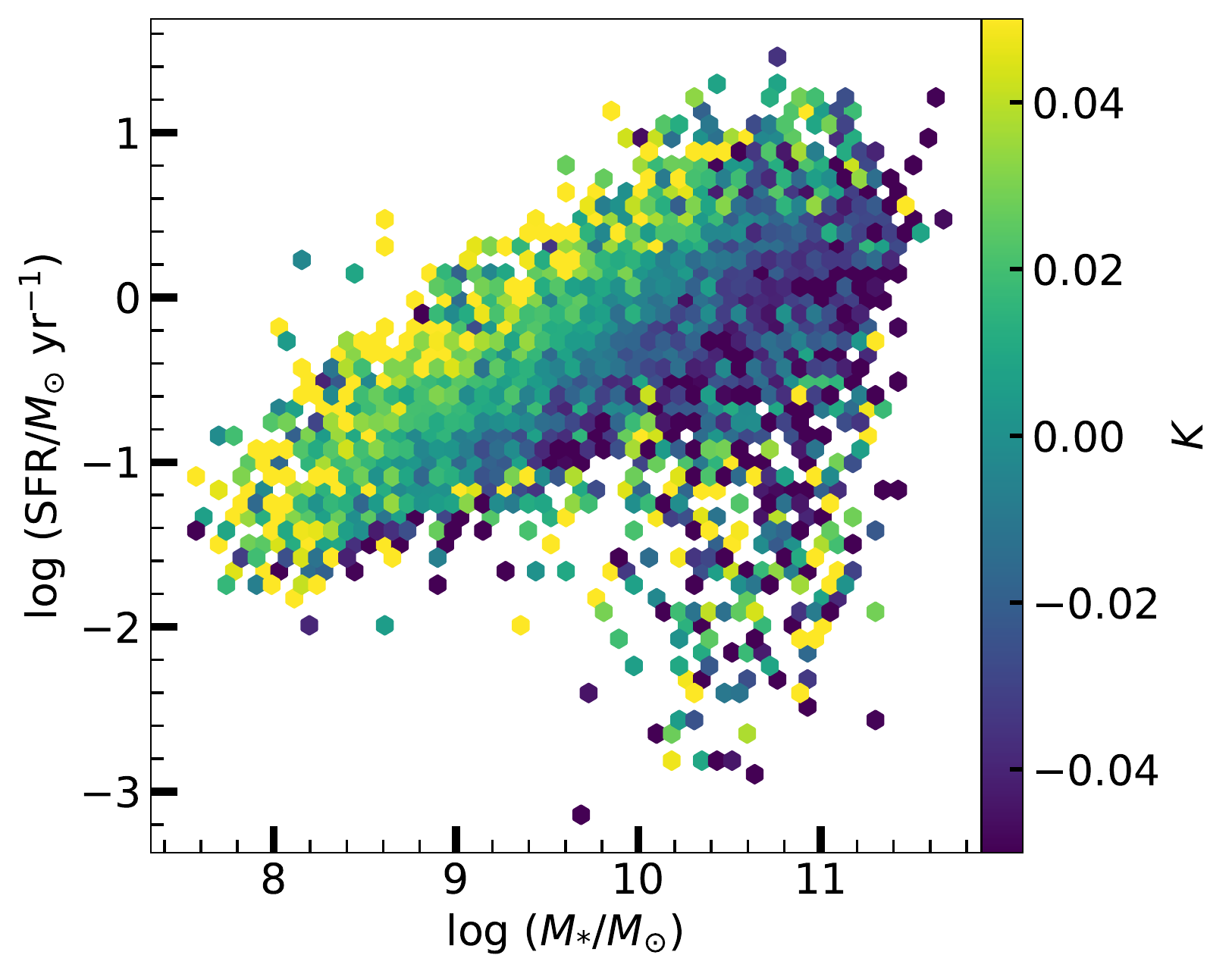}
\caption{The distribution of profile shape $K$ in the space of SFR versus $M_*$. Each pixel is color-coded by the median value of $K$.}
\label{fig:sfr-Con}
\end{figure}

\subsection{Inferring Gas Distribution from \HI\ Profile Shape}
\label{subsec:method}

Modulo the inclination angle along the line-of-sight, the shape of an integrated \HI\, spectrum primarily reflects the spatial and velocity distribution of the neutral atomic hydrogen of the galaxy.  \cite{El-Badry2018MNRAS.477.1536E} highlighted the importance of a rotation or dispersion-dominated velocity field in influencing the observed \HI\ profile shape. However, as recognized by these authors, their simulated low-mass galaxies are on average more dispersion-supported than real galaxies; they have smaller \HI\ radii ($\sim 7$ kpc) and higher \HI\ velocity dispersion ($\sim 30$ \kms) than observed.  Galaxies have global \HI\ velocity dispersions of $\sim$ 10 \kms\ (\citealt{Leroy2008AJ....136.2782L, Ianjamasimanana2015AJ....150...47I, Mogotsi2016AJ....151...15M}), which generally decrease with radius \citep{Tamburro2009AJ....137.4424T}.  \HI\ radii vary from a few to 120 kpc (e.g., \citealt{Wang2016MNRAS.460.2143W}). 

Assuming that the gas and stars are coplanar, we can estimate the inclination angle of the \HI\ disk through the axial ratio of the starlight \citep{Hubble1926ApJ....64..321H}.  Therefore, the main challenge in extracting information on the gas distribution hinges on whether we can constrain the shape of the rotation curve, which depends on both the total mass and the mass distribution of the galaxy (\citealt{Swaters2009A&A...493..871S, Oh2011AJ....142...24O, Lelli2016AJ....152..157L, Sofue2017PASJ...69R...1S, Tiley2019MNRAS.485..934T, Lang2020ApJ...897..122L}).  In relatively massive galaxies, the radial extent amenable to kinematic observations is dominated mostly by the gravitational potential of the stars \citep{Cappellari2016ARA&A..54..597C}, such that, to first order, the rotation curve largely depends on the bulge-to-disk ratio or degree of central concentration of the stellar distribution \citep{Sofue2001ARA&A..39..137S}.  For example, the rotation curve of intermediate-mass, disk-dominated galaxies increases to a maximum and then flattens (e.g., \citealt{Sofue1999PASJ...51..737S}), while in massive, bulge-dominated systems, the rotation curve reaches a maximum more quickly before it flattens, or even declines toward large radii (e.g., \citealt{Noordermeer2007MNRAS.376.1513N}).  By contrast, in low-mass ($M_* \approx 10^7-10^9 \, M_{\mathrm{\odot}}$), late-type galaxies, which exhibit continuously rising rotation curves  (e.g., \citealt{Swaters2009A&A...493..871S, Oh2011AJ....141..193O}), the total mass plays a more important role because the mass distribution of dwarf galaxies is dominated by their dark matter halo at all radii (e.g., \citealt{Oh2011AJ....142...24O, Oh2011AJ....141..193O}).  Since the halo concentration varies with halo mass \citep{Dutton2014MNRAS.441.3359D, Shan2017ApJ...840..104S}, and halo mass correlates tightly with stellar mass \citep{Moster2010ApJ...710..903M}, we can use the stellar mass to control for the effect of halo concentration.  Apart from stellar mass and optical concentration, we note that the size of the \HI\ disk obviously also impacts the radial extent to which the velocity field can be probed.  We use the total \HI\ content to estimate the size of the \HI\ disk (e.g., \citealt{Broeils1997A&A...324..877B, Wang2016MNRAS.460.2143W}).  In the following analysis, we infer the spatial distribution of \HI\ by constructing carefully controlled subsamples of galaxies with fixed $i$, \mstar, $C$, and \MHI.

Figure~\ref{fig:toyModel} schematically sketches how variations in the shape of the rotation curve $V(R)$, the radial distribution of the \HI\ mass surface density $\Sigma_{\rm H\ I} (R)$, and the radial extent of the \HI\ distribution influence the \HI\, line profile and thus $K$.  Ten examples of different combinations of $V(R)$ and $\Sigma_{\rm H\ I}(R)$ are shown, for a fixed inclination angle.  At a given centrally peaked \HI\ radial distribution (row a), the rising rotation curve typical of low-mass, dwarf irregular galaxies produces a narrow single-peaked profile and a large, positive value of $K$ (blue), as commonly observed (\citealt{Oh2011AJ....141..193O, Hunter2021AJ....161...71H}).  By contrast, an initially rising and then flattened rotation curve of more massive galaxies generates a broader single-peaked profile, with lower but still positive $K$ (black). A more steeply rising and then flattened rotation curve yields a double-horned profile and $K < 0$ (green).  If, instead, we fix the rotation curve to that typical of spiral galaxies (row b), \HI\ radial distributions of different degrees of central concentration correspond to either a single-peaked (blue) or flat-topped (black) profile with $K \gtrsim 0$, a flat $\Sigma_{\rm H\ I}(R)$ distribution produces a typical double-horned profile with negative $K$ (green), and a central hole generates the most prominent double horns and the most negative values of $K$ (red).  The central \HI\ hole in some spiral galaxies can have a radius of a few to 20 kpc \citep{Murugeshan2019MNRAS.483.2398M}.  Finally, for a given rotation curve and flat $\Sigma_{\rm H\ I}(R)$ (row c), galaxies with larger \HI\ radii (thus higher \HI\ content) have more pronounced double-horned profiles and smaller $K$. 

\begin{figure*}[ht]
\epsscale{1.}
\plotone{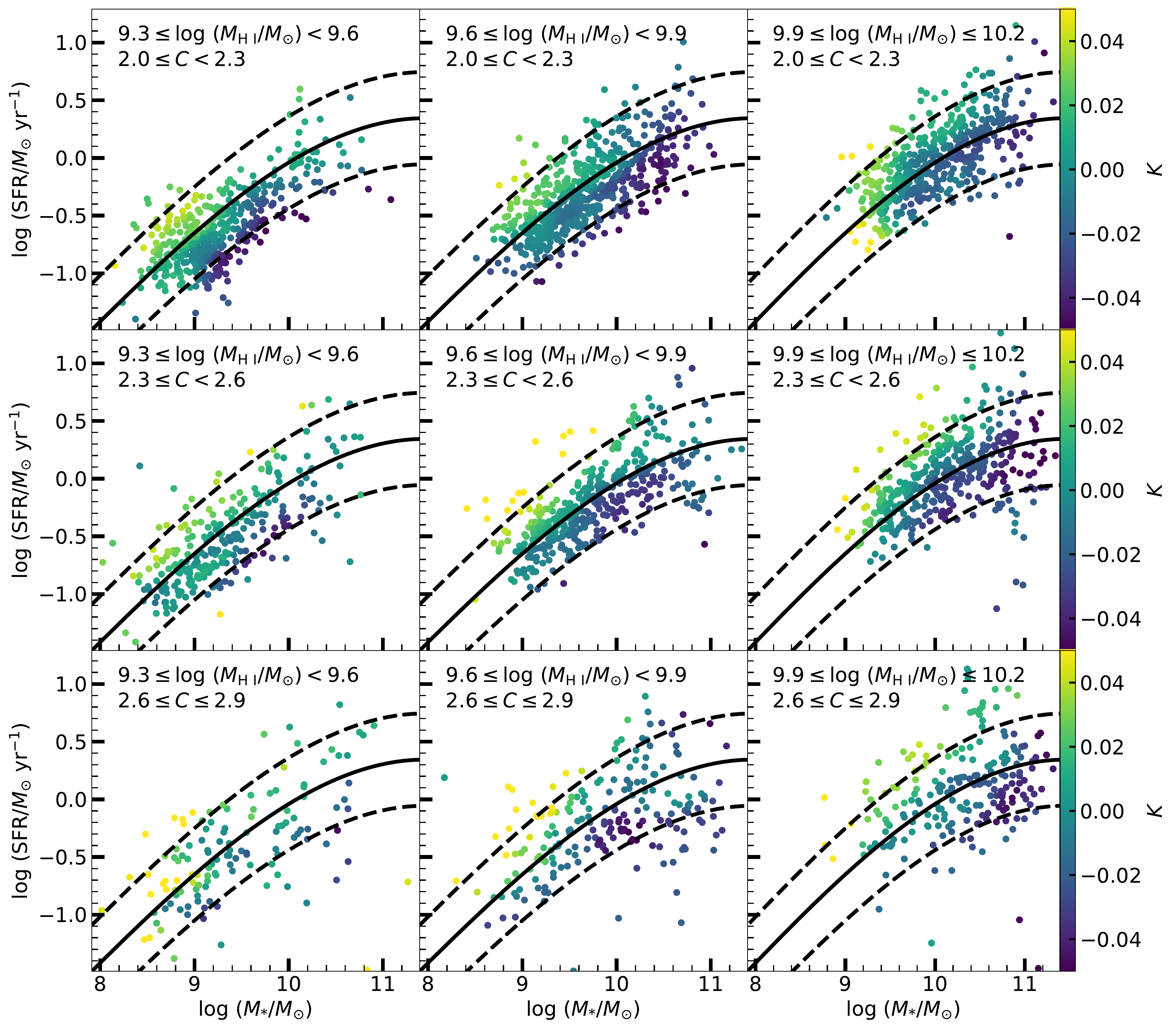}
\caption{The distribution of $K$ in the space of SFR versus \mstar, for galaxies with $9.3 \leq \log (M_{\mathrm{H~I}}/M_\odot) < 9.6$ (left), $9.6\leq\log (M_{\mathrm{H~I}}/M_\odot) <9.9$ (middle), and $9.9 \leq\log (M_{\mathrm{H~I}}/M_\odot) \leq 10.2$ (right).  From top to bottom, the optical concentration for each row is $2.0 \leq C< 2.3$, $2.3 \leq C< 2.6$, and $2.6 \leq C\leq 2.9$. The black solid and dashed curves mark the mean position and $\pm$ 0.4 dex of the star-forming galaxy main sequence \citep{Saintonge2016MNRAS.462.1749S}.  The inclination angle is fixed to 50$^{\circ}\leq i< 70^{\circ}$. The color of the data points is the LOESS-smoothed value of $K$. }
\label{fig:mass-sfr-scog}
\end{figure*}

\begin{figure*}
\epsscale{1.0}
\plotone{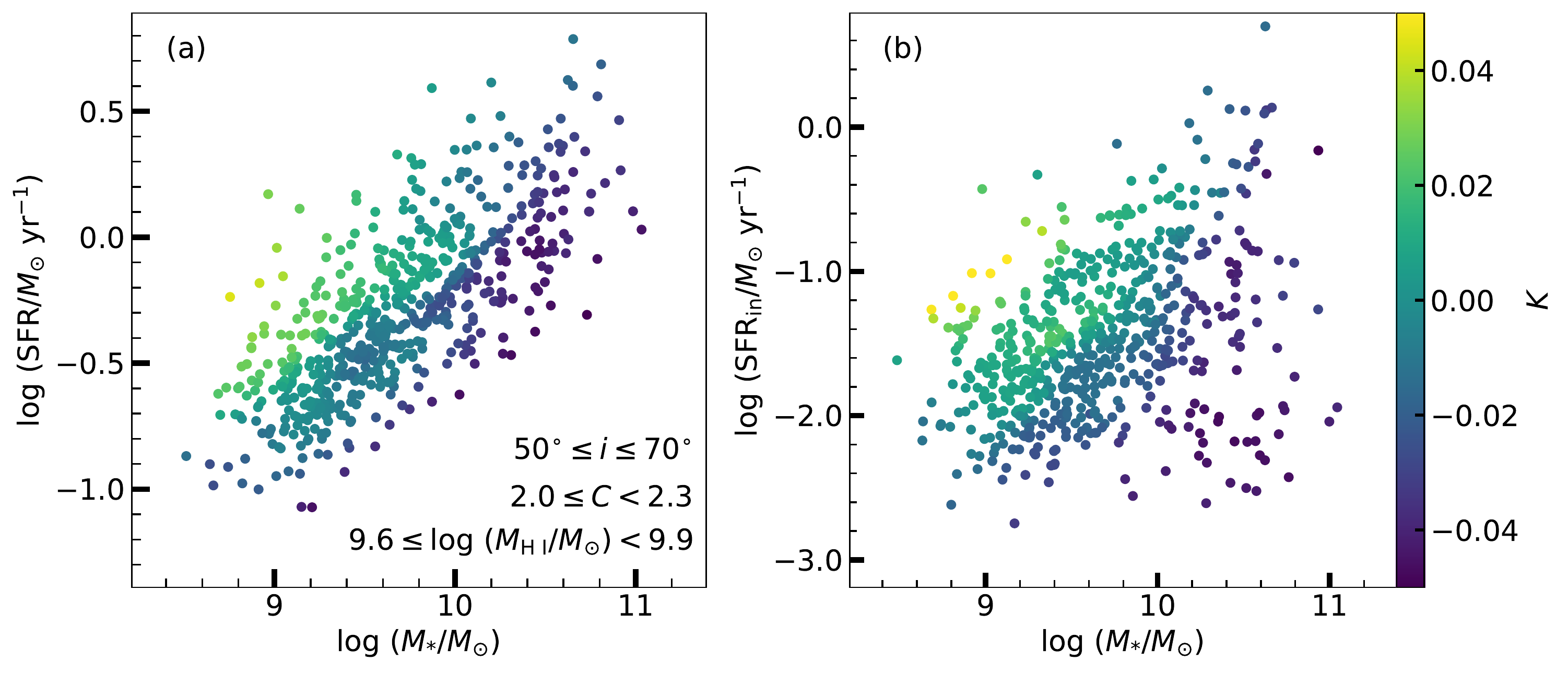}
\caption{The distribution of $K$ in the space of (a) SFR versus \mstar\ and (b) SFR$_{\rm in}$ versus $M_*$, for galaxies with $i$, $C$, and \MHI\ fixed to the values given in the legend of panel (a). The color of the data points is the LOESS-smoothed value of $K$.}
\label{fig:mass-sfrin-kappa}
\end{figure*}

\section{Relation between SFR and \HI\ Distribution}
\label{sec:HISFR}

Figure~\ref{fig:sfr-Con} clearly shows that $K$ increases with increasing SFR at a given \mstar.  
It is tempting to conclude from this trend that the \HI\ radial distribution is associated with the level of star formation in a galaxy. However, as discussed in Section~\ref{subsec:method}, $K$ can be influenced by several factors, chief among them the inclination angle, stellar mass, optical concentration, and \HI\ mass.  These parameters must be properly controlled before inferences can be made concerning the spatial distribution of the \HI.

As an example of our procedure, we select galaxies with 50$^{\circ}\leq i< 70^{\circ}$ and further divide them into subsamples in narrow bins of $C$ and \MHI, to study the distribution of $K$ along the star-forming galaxy main sequence (Figure~\ref{fig:mass-sfr-scog}). The data points are LOESS-smoothed following \citet{Cappellari2013MNRAS.432.1709C}.  Defining the star formation rate deviation ($\Delta \log \rm SFR$; Appendix~\ref{appsec:delta}) as the vertical offset from the main sequence \citep{Saintonge2016MNRAS.462.1749S}, Table~\ref{tab:pearson} summarizes the correlation analysis between $K$ and $\Delta \log \rm SFR$. We caution that the scatter around the star-forming main sequence is partly induced by uncertainties in the measurements of stellar mass and SFR, which may smooth out and obscure weak physical trends from our analysis. More than half (15/28) of the galaxy subsamples show a statistically significant correlation between $K$ and $\Delta \log \rm SFR$ (probability $p < 0.05$ of rejecting the null hypothesis that the two parameters are uncorrelated according to a Pearson correlation analysis).  We obtain similar trends for subsamples with 30$^{\circ}\leq i< 50^{\circ}$ or 70$^{\circ}\leq i\leq\ 90^{\circ}$.

We interpret the systematic variation of $\Delta \log \rm SFR$ with $K$ as an indication that the star formation activity of galaxies is regulated, at least in part, by the location of the \HI\ gas within the galaxy.  Galaxies with larger values of $K$, indicative of more centrally concentrated \HI\ distributions, display systematically higher SFR at fixed \mstar. This result qualitatively supports the notion that increasing the gas concentration in a galaxy elevates its SFR (e.g., \citealt{Hopkins2011MNRAS.415.1027H, Tacchella2016MNRAS.457.2790T}), pointing to gas concentration as an important physical parameter responsible for the observed scatter of the main sequence.  While stars form from molecular, not atomic\footnote{Except possibly under extreme conditions in the early Universe, when neutral atomic hydrogen may directly collapse into stars (e.g., \citealt{Bromm2011ARA&A..49..373B}).}, gas, the conversion of \HI\ to ${\rm H}_2$ requires a minimum critical total gas mass surface density (\citealt{Bigiel2008AJ....136.2846B, Leroy2008AJ....136.2782L}) that is most effectively achieved within the optical disk.  Under ordinary circumstances, the majority of the \HI\ content in spiral galaxies outside of clusters accumulates at large radii (\citealt{Kreckel2012AJ....144...16K}). In order for molecular gas, and hence stars, to form, the \HI\ must migrate inward into the higher density environment of the optical disk.  Whatever the mechanism may be that governs the radial transport of the \HI\ (for a recent discussion, see \citealt{Yesuf2021ApJ...923..205Y}), our results demonstrate the importance of the spatial location of the atomic gas, which mediates  the variation of star formation activity in galaxies by facilitating the production of molecular gas.
\citet{Wang2020ApJ...890...63W} reported broadly consistent results, finding that the SFR at a given stellar mass correlates well with the average \HI\ surface density within the optical radius of the galaxy. Their work assumed an average \HI\ radial distribution and was limited solely to disk-dominated galaxies to mitigate the large dispersion in the \HI\ radial profiles of bulge-dominated systems \citep{Wang2016MNRAS.460.2143W}. The $K$ method presented in this paper obviates the need to assume any specific \HI\ radial distribution, and it can be applied to galaxies of all morphological types. 

Where within the galaxy does the star formation enhancement actually occur?  The natural candidate is the central region of the galaxy \citep{Ellison2018MNRAS.474.2039E}, whose high gas and stellar mass surface density are conducive to efficient star formation \citep{Leroy2008AJ....136.2782L}.  To test this supposition, we repeat the analysis by replacing the total, globally averaged SFR (Figure~\ref{fig:mass-sfrin-kappa}a) with the central SFR (SFR$_{\rm in}$) as measured within the 3$^{\prime\prime}$-diameter SDSS fiber (Figure~\ref{fig:mass-sfrin-kappa}b), which encompasses a physical scale of $\sim 2$~kpc at a median sample redshift of 0.033. As with the total SFR, SFR$_{\rm in}$ increases with increasing $K$.  Calculating $\Delta \log {\rm SFR}_{\rm in}$ as the vertical deviation from the star-forming main sequence defined using SFR$_{\rm in}$ (Appendix~\ref{appsec:delta}; Figure~\ref{fig:deltaSFR}b), the global trend between $K$ and offset from the star-forming main sequence is preserved, as is its statistical significance (Table~\ref{tab:pearson}).  Even though the enhancement of inner SFR is similar to that of global SFR, the systematical uncertainties of SFR$_{\rm in}$ are non-negligible. In our sample SFR$_{\rm in}$, limited to a single, 3$^{\prime \prime}$ aperture,  accounts for only $\sim 10\%$ of the total SFR.  At higher redshifts, this fixed aperture probes a larger physical size than that of the same galaxy at lower redshifts. At the same redshift, the fixed aperture also measures a larger physical scale with respect to the scale of the whole stellar disk of a low-mass galaxies than for a high-mass galaxies.  We assess the impact of this aperture effect by restricting the subsamples further to narrow ranges in redshift ($\Delta z< 0.01$).  The four subsets with more than 100 galaxies that remain continue to support a significant correlation between $K$ and $\Delta \log {\rm SFR}_{\rm in}$.

The total reservoir of cold baryons available to fuel star formation in galaxies resides predominantly in the form of neutral atomic hydrogen gas, which resides mostly on scales larger than the optical disk.  Star formation can transpire efficiently only after gas moves or is transported from the outer disk to the inner disk, where the \HI\ surface density can exceed the threshold for H$_2$ to form ($\sim 10$ \msun\, pc$^{-2}$; \citealt{Krumholz2015MNRAS.453..739K, Krumholz2017MNRAS.466.1213K}). In support of this picture, \cite{Wang2020ApJ...890...63W} showed that the \HI\, surface density within the optical radius is critical for star formation in galaxies.  The angular momentum of the gas can be redistributed by internal processes due to bars (e.g., \citealt{Fanali2015MNRAS.454.3641F}) and spiral arms (e.g., \citealt{YuSY2021ApJ...917...88Y, YuSY2022}), and external processes such as minor mergers (e.g., \citealt{Mihos1994ApJ...425L..13M}), major merger (e.g., \citealt{Toomre1972ApJ...178..623T, Barnes2004MNRAS.350..798B, Larson2016ApJ...825..128L}), and tidal interactions (e.g., \citealt{Larson2002MNRAS.332..155L}).  

We cannot readily assess which, if any, of these mechanisms is primarily responsible for shaping the gas distribution of the galaxies in our sample, although we note that major mergers are unlikely to contribute significantly, both because their incidence is low in the local Universe ($\sim 2\%$; \citealt{Patton2008ApJ...685..235P}) and because our procedure for identifying the optical counterpart to each ALFALFA source deliberately excludes close galaxy pairs (see details in \citealt{Yu2022}).

\bigskip
\section{Conclusions}
\label{sec:sum}

We use a new, comprehensive catalog of \HI\ profile measurements of over 13,000 nearby galaxies that overlap between ALFALFA and SDSS to investigate the influence of \HI\ spatial distribution on star formation activity.  Building upon the curve-of-growth method of \cite{Yu2020ApJ...898..102Y}, \citet{Yu2022} introduce a new parameter $K$ to describe the observed \HI\ profile shape. We extract statistical information on the spatial distribution of neutral atomic hydrogen based solely on the shape of the integrated \HI\ 21~cm profile of nearby galaxies.  For galaxies of fixed inclination angle, \mstar, $C$, and \MHI, centrally peaked \HI\ distributions produce single-peaked lines ($K > 0$), while a centrally flat or depressed \HI\ profile leads to flat or double-horned line shapes ($K \leq 0$).  

We provide the first qualitative, statistically robust evidence that the global concentration of \HI\ is linked with the scatter around the star-forming galaxy main sequence.  The deviation of a galaxy from the star-forming main sequence correlates systematically with $K$.  No strong differences are observed when global SFRs are replaced with SFRs within the central $\sim 2$ kpc, suggesting that the \HI\ distribution influences star formation throughout the optical disk.
We interpret these trends to signify that higher central concentration of \HI\ within the optical disk facilitates the conversion of neutral atomic hydrogen to molecular hydrogen, a prerequisite for star formation.  These results highlight the importance of understanding the physical location of the baryonic reservoir of \HI\ and the mechanisms by which it migrates inward to fuel star formation.

Our sample, based on the ALFALFA survey, is biased against quiescent galaxies \citep{Huang2012ApJ...756..113H}.  Our conclusions may not hold for galaxies with low SFR or low \HI\, content.  It would be fruitful to extend this work using future deeper blind \HI\ surveys, as well as to attempt a stacking analysis of undetected sources from ALFALFA itself.  Single-dish, global spectra are, of course, no substitute for resolved maps. The actual spatial distribution of \HI\ is often quite complex, exhibiting warps (\citealt{Sancisi1976A&A....53..159S, Martinsson2016AA...585A..99M}) and asymmetric features \citep{Richter1994A&A...290L...9R, Haynes1998AJ....115...62H, Watts2020MNRAS.492.3672W} that may be poorly captured in an integrated spectrum. In extreme cases, such as very low-mass dwarf irregular galaxies whose gas dispersion can be similar to its level of rotation, a large \HI\ velocity dispersion can affect the profile shape \citep{El-Badry2018MNRAS.477.1536E}. Our analysis assumes that the \HI\ velocity dispersion is low with respect to the rotational velocity at all radii.  We look forward to future insights into the spatial distribution of \HI\ in galaxies from surveys by the Square Kilometer Array (\citealt{Dewdney2009IEEEP..97.1482D}) and its pathfinders  (ASKAP: \citealt{Johnston2007PASA...24..174J}; MeerKAT: \citealt{Jonas2009IEEEP..97.1522J}; Murchison Widefield Array: \citealt{Tingay2013PASA...30....7T}).

\acknowledgments

We thank the anonymous referee and statistics editor for helpful comments and suggestions. This work was supported by the National Science Foundation of China (11721303, 11991052, 11903003, 12073002), the China Manned Space Project (CMS-CSST-2021-A04, CMS-CSST-2021-B02), and the National Key R\&D Program of China (2016YFA0400702).  We are grateful to Martha Haynes for kindly providing the spectra of ALFALFA.  YNK thanks Pei Zuo, Yuming Fu, Tianqi Huang, Bitao Wang, and Jinyi Shangguan for useful advice and discussions. This research made use of the NASA/IPAC Extragalactic Database ({\url http://ned.ipac.caltech.edu}), which is funded by the National Aeronautics and Space Administration and operated by the California Institute of Technology. We used Astropy, a community-developed core Python package for astronomy \citet{AstropyCollaboration2013A&A...558A..33A}.

\appendix
\section{The Vertical Deviations of SFR and SFR$_{\rm in}$}
\label{appsec:delta}

\begin{figure*}[t]
\epsscale{1.0}
\plotone{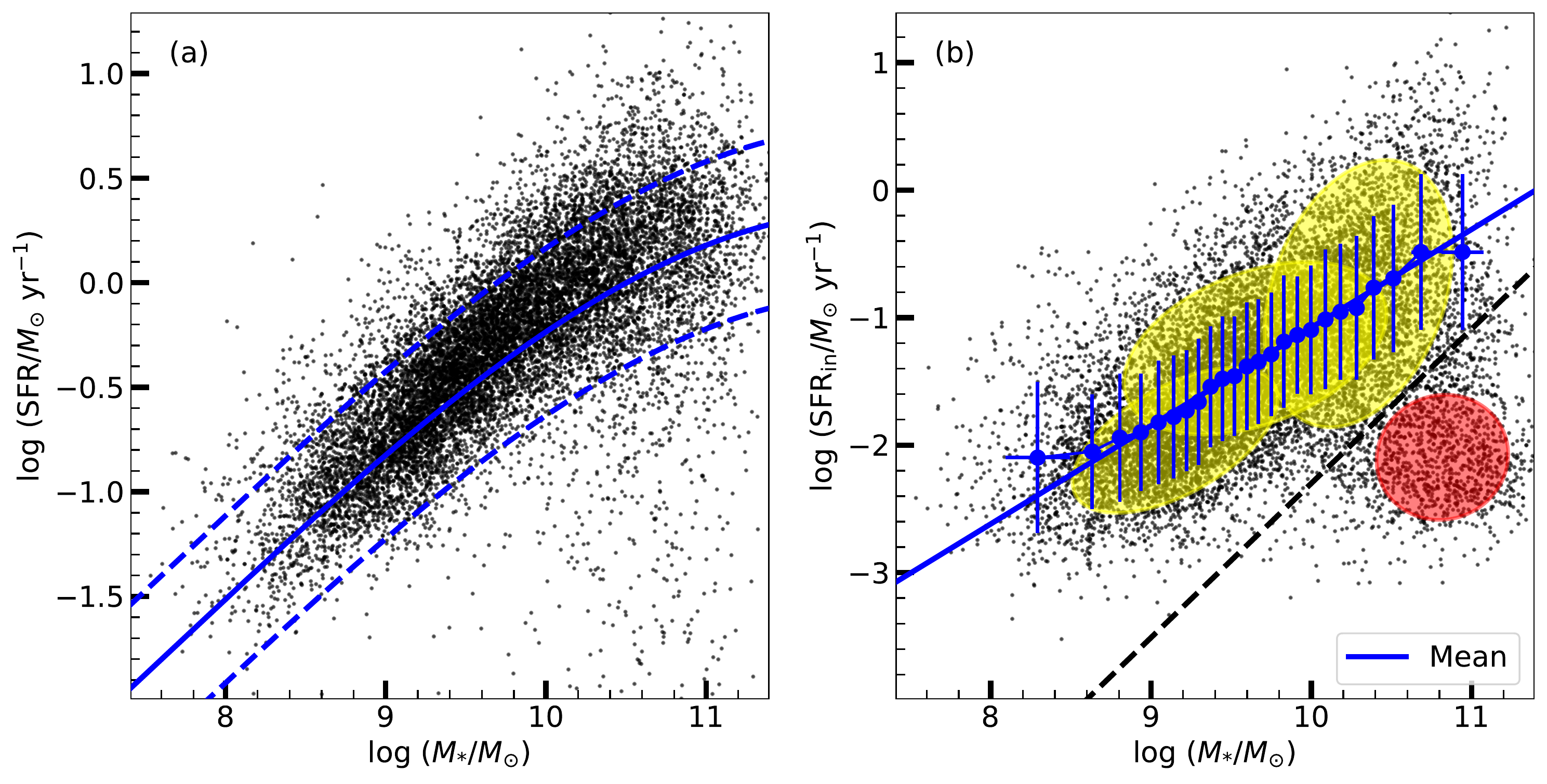}
\caption{The relation between \mstar\ and (a) SFR and (b) SFR$_{\rm in}$. In panel (a), the blue solid and dashed curves mark the mean position and $\pm$ 0.4 dex of the star-forming galaxy main sequence from \cite{Saintonge2016MNRAS.462.1749S}. In panel (b), the blue line is a linear fit to the mean values in each bin (blue points) after excluding galaxies below the dashed black line. The yellow and red region show the four components of the Gaussian mixture model.}
\label{fig:deltaSFR}
\end{figure*}

We define the global star formation rate deviation ($\Delta \log \rm SFR$) as the vertical deviation to the relation between log~SFR and log~\mstar\ (Figure~\ref{fig:deltaSFR}a; \citealt{Saintonge2016MNRAS.462.1749S}). Following \citet{FraleyRaftery2002}, we fit the data points in Figure~\ref{fig:deltaSFR}b with a four-component Gaussian mixture model. Based on the Gaussian mixture components, we exclude galaxies lying below the black dashed line, $\log\ \rm{SFR}_{\rm in} = 1.21\log M_*  - 14.4$, to select centrally star-forming galaxies. We fit a linear trend, following \citet{Cappellari2013MNRAS.432.1709C}, to mean values of $\log\ \rm{SFR}_{\rm in}$ in bins of $\log M_*$:

\begin{equation}
\log\ \rm{SFR}_{\rm in} = 0.77\log M_*  - 8.8.
\label{equ:deltaSFRin}
\end{equation}

\noindent
The central star formation rate deviation ($\Delta \log {\rm SFR}_{\rm in}$) is the vertical deviation to {\bf Equation (A1)}.

\vfill\eject


\end{CJK*}
\end{document}